\documentclass[twocolumn,tighten]{aastex63}
\usepackage{ulem}

\usepackage{amsmath}
\usepackage{multirow}

\usepackage{lineno}

\begin{document}

\title{Testing photoevaporation and MHD disk wind models through future high-angular resolution radio observations:
the case of TW Hydrae}

\author{Luca Ricci}
\affiliation{Department of Physics and Astronomy, California State University Northridge, 18111 Nordhoff Street, Northridge, CA 91330, USA}

\author{Sarah K. Harter}
\affiliation{Department of Physics and Astronomy, California State University Northridge, 18111 Nordhoff Street, Northridge, CA 91330, USA}

\author{Barbara Ercolano}
\affiliation{University Observatory, Faculty of Physics, Ludwig-Maximilians-Universität München, Scheinerstr. 1, 81679 Munich,
Germany}

\author{Michael Weber}
\affiliation{University Observatory, Faculty of Physics, Ludwig-Maximilians-Universität München, Scheinerstr. 1, 81679 Munich,
Germany}

\begin{abstract}
We present theoretical predictions for the free-free emission at cm wavelengths obtained from photoevaporation and MHD wind disk models adjusted to the case of the TW Hydrae young stellar object. For this system, disk photoevaporation with heating due to the high-energy photons from the star has been proposed as a possible mechanism to open the gap observed in the dust emission with ALMA. We show that the photoevaporation disk model predicts a radial profile for the free-free emission that is made of two main spatial components, one originated from the bound disk atmosphere at 0.5-1 au from the star, and another more extended component from the photoevaporative wind at larger disk radii. We also show that the stellar X-ray luminosity has a significant impact on both these components. The predicted radio emission from the MHD wind model has a smoother radial distribution which extends to closer distances to the star than the photoevaporation case. We also show that a future radio telescope such as the \textit{Next Generation Very Large Array} (ngVLA) would have enough sensitivity and angular resolution to spatially resolve the main structures predicted by these models.

\end{abstract}

\section{Introduction}

Planets form from the gas and dust particles contained in disks orbiting young stars. The overall time evolution and final dispersal of these planet-forming disks play a major role in determining the properties of the planets born out of these systems. 
Hence, understanding the physical processes which regulate the late evolution and final dispersal of young disks is key for any theory of planet formation \citep[e.g.,][]{Alexander:2014, Ercolano:2017}.  

The X-ray photoevaporation model \citep{Ercolano:2008,Owen:2010, Owen:2011,Owen:2012, Picogna:2019} has been successful in explaining the observed two-timescale and inside-out dispersal of most disks \citep{Luhman:2010, Koepferl:2013, Ercolano:2017}. In this model, soft X-rays (0.1 keV $< E <$  1 keV) from the central star can penetrate the disk atmosphere and efficiently heat up the disk. In the regions of the disk in which the sound speed of the heated gas exceeds the local escape velocity due to stellar gravity, gas flows away from the disk, and a thermal wind is centrifugally launched \citep[e.g.,][]{Owen:2010}. When the mass loss rate due to the wind and the mass accretion rate become comparable, a gap forms in the disk, followed by an inner hole which quickly grows in stellocentric radius, with the outer disk regions being dispersed from the inside-out.

Although these models can reproduce the spectral profile of emission lines which are thought to originate in the disk wind \citep[e.g.,][]{Ercolano:2010, Ercolano:2016,Weber:2020}, no conclusive evidence for a gap opened by photoevaporation has been obtained yet. 
Interestingly, \citet{Ercolano:2017b} have shown that models of X-ray photoevaporation can explain the gap observed at a stellocentric radius of 1 au in the dust emission of the TW Hydrae disk \citep{Andrews:2016} with the \textit{Atacama Large Millimeter/submillimeter Array} (ALMA), together with other properties of that system (e.g., mass accretion rate).

As shown by \citet{Owen:2013}, hydrodynamic models of disk photoevaporation coupled to radiative transfer post-processing techniques predict free-free emission produced from a hot bound X-ray heated atmosphere and/or from the photoevaporation wind with specific radial substructures \citep[see also][]{Pascucci:2014}. In this work, we produce similar predictions for the free-free radio emission from the X-ray photoevaporation models presented in \citet{Ercolano:2017b} adjusted to the case of the TW Hydrae (TW Hya) disk. 

A separate mechanism that has been proposed to contribute to the disk dispersal, as well as to play a role in the redistribution of angular momentum in the disk, are magneto-hydrodynamic (MHD) winds \citep{Blandford:1982,Anderson:2003,Bai:2016}. Contrary to the purely hydrodynamic thermal winds driven by photoevaporation, the efficiency of MHD winds strongly depends on the net magnetic flux retained in mature disks. Although the evolution of MHD winds depend on a number of assumptions, some qualitative differences with the photoevaporation winds have been identified in the literature. For example, gas in MHD winds can be accelerated also at very small stellocentric radii where photoevaporation winds are inefficient because of the strong gravitational field of the star. Hence, it is interesting to investigate whether these differences can be traced and tested also through the radio emission of protoplanetary disks, which can be imaged at high angular resolution using radio interferometry. 

In this paper we follow the procedure outlined in \citet{Weber:2020} to calculate the radio emission expected from a MHD wind adapted to the case of the TW Hya system, and compare these predictions to the expected radio emission from models of X-ray driven photoevaporation.

We also show that a future radio interferometer like the Next Generation Very Large Array \citep[ngVLA,][]{Murphy:2018} would be capable of detecting and spatially resolving several of the predicted substructures in the free-free emission at cm wavelengths from the photoevaporation and MHD winds models. The comparison with the theoretical predictions would provide key constraints to some important parameters for models of disk dispersal, together with conclusive evidence for the ability of either photoevaporation or MHD winds to produce gaps and other substructures in the morphology of young disks. 

Section~\ref{sec:methods} describes the main properties of the X-ray photoevaporation and MHD wind models adapted to the case of TW Hya, and the methods used to derive the model synthetic images at a wavelength of 1 cm. 
Section~\ref{sec:results} shows the results of these calculations, as well as predictions of future ngVLA observations to detect and spatially resolve the expected disk substructures in the free-free radio emission. Sections~\ref{sec:discussion} and~\ref{sec:conclusions} outline the discussion and conclusions of this work, respectively.

\section{Methods: from disk simulations to ngVLA observations}
\label{sec:methods}

In this Section we present the methods used to estimate the radio free-free emission from ionized gas in the two contests of X-ray photoevaporation and MHD disk models applied to the TW Hya disk. We then describe how we used the results of these calculations to derive predictions for future radio observations with the ngVLA. We start by summarizing the main properties of the TW Hya system which are relevant for this work.

\subsection{The case of TW Hydrae}
\label{sec:TW}

TW Hya is the closest gas-rich young disk to Earth \citep[distance $\approx$ 60.1 pc,][]{GAIA:2018}. With a stellar mass of $\approx 0.8-0.9~M_{\odot}$ \citep{Huang:2018}, it is a member of the TW Hya association, which has a relatively advanced age of $\sim 10$ Myr \citep[e.g.,][]{Weinberger:2013}. Together with its nearly face-on orientation in the sky \citep[disk inclination $\approx$ 7 deg,][]{Qi:2004,Andrews:2016}, these properties make this system a unique laboratory to investigate at the best possible spatial resolution the imprints on the disk structure left by the physical mechanisms which are responsible for the disk evolution and dispersal. 

Several hi-res optical observations \citep[e.g.,][]{Debes:2017} and interferometric observations at mid-IR \citep[e.g.,][]{Ratzka:2007} and sub-mm/mm wavelengths \citep[e.g.,][]{Andrews:2016} have resolved a variety of different substructures at different stellocentric radii.
In particular, significant depletion of dust within the first couple of au from the star has been inferred from the analysis presented in \citet{Calvet:2002,Eisner:2006,Hughes:2007,Menu:2014}, as well as from the detection of a gap at 1 au from the star via direct imaging of the dust continuum emission with ALMA at 0.87 mm \citep{Andrews:2016}. 

The low values inferred for the mass accretion rate of the TW Hya star \citep[$\sim 4 \times 10^{-10} - 2 \times 10^{-9} ~M_{\odot}~\rm{yr}^{-1}$, ][]{Muzerolle:2000,Brickhouse:2012,Manara:2014} and its relatively high X-ray luminosity ($\sim 2 \times 10^{30}$ erg s$^{-1}$)   extrapolated in the 0.1 keV $< E <$ 10 keV spectral range \citep{Robrade:2006} make the inner regions of this disk susceptible to significant mass loss by X-ray driven photoevaporation \citep{Owen:2010}. In fact, evidence of ongoing photoevaporation in the TW Hya disk has been presented from the analysis of the [NeII] 12.8 $\micron$ \citep{Herczeg:2007,Pascucci:2009} and [OI]6300 emission lines \citep{Pascucci:2011,Ercolano:2016}.

\citet{Ercolano:2017b} built on these results to show that also the dust depletion observed at mid-IR to sub-mm wavelengths is consistent with the expectations of X-ray photoevaporation disk models, making this the first candidate object where a photoevaporative gap may have been imaged around
the time at which it is being created. 

\subsection{X-ray photoevaporation disk model}
\label{sec:X-ray_model}
We used the methods described by  \citet{Picogna:2019} to obtain radiation-hydrodynamical simulations of an X-ray photoevaporated disk around a young star with properties similar to TW Hya as described by \citet{Ercolano:2017b}. We refer the reader to the cited articles for details about the numerical methods and the input parameters. For the X-ray luminosity $L_X$ in the models, we considered two values of $10^{29}$ and $10^{30}$ erg s$^{-1}$, respectively. Although the latter value is more in line with the observational results for TW Hya (Section~\ref{sec:TW}) we decided to run simulations also with a lower value for an initial exploration of the dependence on $L_X$ of the results from our models.

We then post-processed the grids using an improved version of the {\sc mocassin} radiative transfer code \citep{Ercolano:2003, Ercolano:2005, Ercolano:2008} to obtain images at a wavelength of 1~cm. 
In Appendix A we describe the method used in this work to reduce the Monte Carlo noise at long wavelengths.


\subsection{MHD wind model}
\label{sec:MHD_model}

In order to estimate the spatial distribution of the free-free emission from a MHD wind, we used the simple magnetocentrifugally driven wind model presented in \citet{Weber:2020}. This is based on an analytical description with the density and velocity structure obtained by \citet{Milliner:2019} from the \citet{Blandford:1982} axisymmetric self-similar solutions for a magnetocentrifugally driven MHD wind for thin disks. More details can be found in \citet{Milliner:2019} and \citet{Weber:2020}. In particular, the model considered here is model \textit{MHD-1} in \citet{Weber:2020}, with a mass-loss rate of $10^{-8.6}~M_{\odot}/$yr and an X-ray luminosity for the star of $L_X = 2 \times 10^{30}$ erg s$^{-1}$. 

Like for the X-ray photoevaporation disk model presented in the last section, a synthetic image at a wavelength of 1 cm was obtained using the {\sc mocassin} code. 

\subsection{Simulations of the ngVLA observations}
\label{sec:ngvla}

The model images obtained in Sections~\ref{sec:X-ray_model} and \ref{sec:MHD_model} were converted into predictions for future observations with the ngVLA at $\lambda =$ 1 cm using the CASA software package \citep{McMullin:2007}.

Given the expected angular resolution ($\approx$ milliarcsec) and sensitivity at cm wavelengths, a future ngVLA has the potential to spatially resolve structures in the ionized gas emission in disks in nearby star forming regions, as expected by models of disk photoevaporation and MHD disk winds. 

To simulate the results of future interferometric observations with the ngVLA, we adopted the same procedure as in \citet{Ricci:2018} and \citet{Harter:2020}, which used the \texttt{SIMOBSERVE} task to generate the visibility dataset in the $(u,v)$ Fourier space, and the \texttt{SIMNOISE} task to add the noise by corrupting the visibilities.

For the ngVLA simulations we considered the original ngVLA Rev B array configuration with antennas
distributed across the US Southwest and Mexico. This configuration includes 214 antennas of 18 meter diameter, with baselines up
to 1000\,km \citep{Selina:2018}. 

For the imaging of the interferometric visibilities we employed the \texttt{CLEAN} algorithm with Briggs weighting, and adjusted the robust parameter to give a reasonable synthesized beam and noise performance.
In particular, the ngVLA images were computed
with a Briggs weighting scheme with robust parameter $R = -2$ (uniform weighting). We also employed a multiscale clean approach to better recover compact emission at both high brightness and larger and more diffuse structures in the model.
The disk was centered at the location of TW Hydrae, i.e. RA(J2000)$=$11:01:51.90, Dec(J2000)$=$-34:42:17.0, and the assumed distance is 60.1 pc \citep[][]{GAIA:2018}.


\begin{figure*}[t!]
\begin{center}
\includegraphics[scale=1.28]{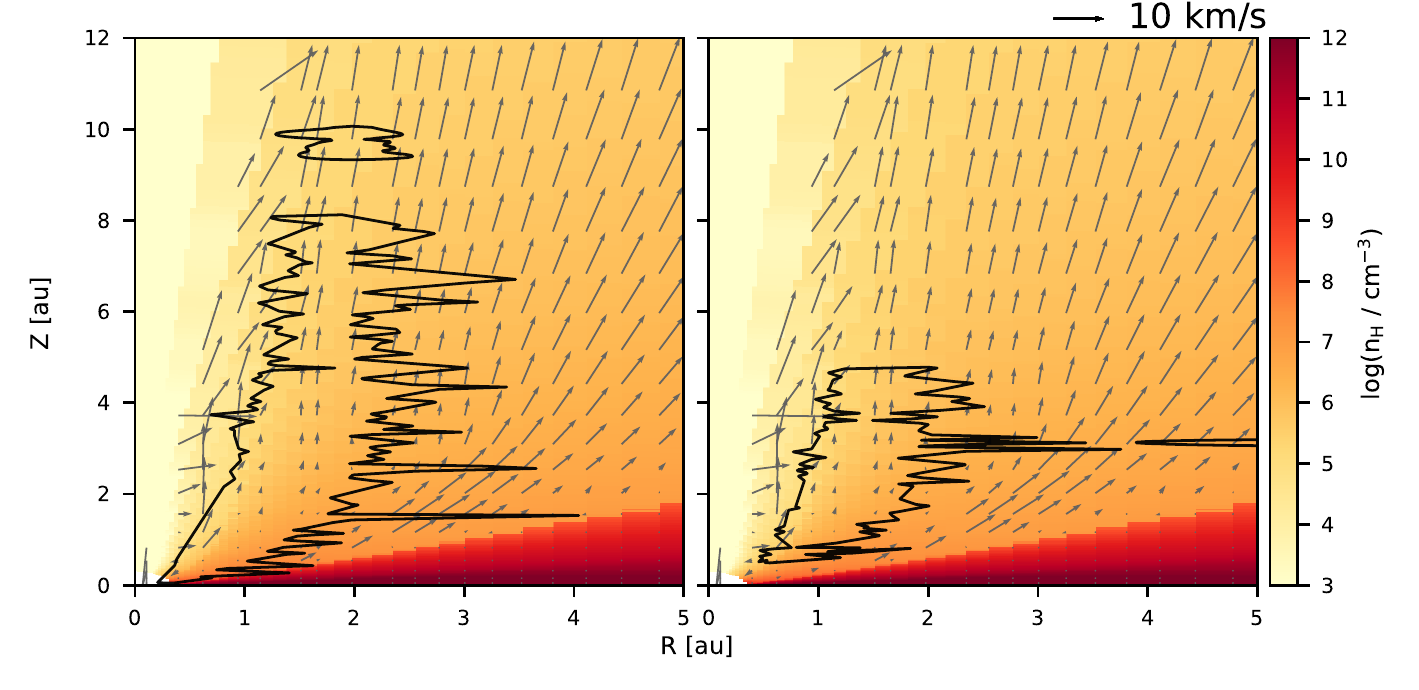}
\end{center}
\caption{Neutral hydrogen density map for the photoevaporation disk models presented in Section~\ref{sec:X-ray_model} as a function of the radial distance from the star and vertical coordinate from the disk midplane. Left and right panels are for models with $L_X = 10^{29}$ erg s$^{-1}$ and $10^{30}$ erg s$^{-1}$, respectively. Vectors indicate the direction of the gas flow and their length represents the speed of the flow with the reference speed of 10 km s$^{-1}$ shown above the top right corner of the right panel (relation between vector length and speed is linear). The superimposed contour lines define the 90\% emission region for the free-free emission at 1 cm, after a 2D gaussian smoothing with $\sigma = 2$ grid cells to make the contour lines less noisy and confusing. Note that in both panels the scaling of the x and y axes are different, and the significant stretch of the y-axis was chosen to better highlight the wind structure in the vertical direction.
}
\label{fig:X-ray_density}
\end{figure*}

\begin{figure*}[t!]
\begin{center}
\includegraphics[scale=1.4]{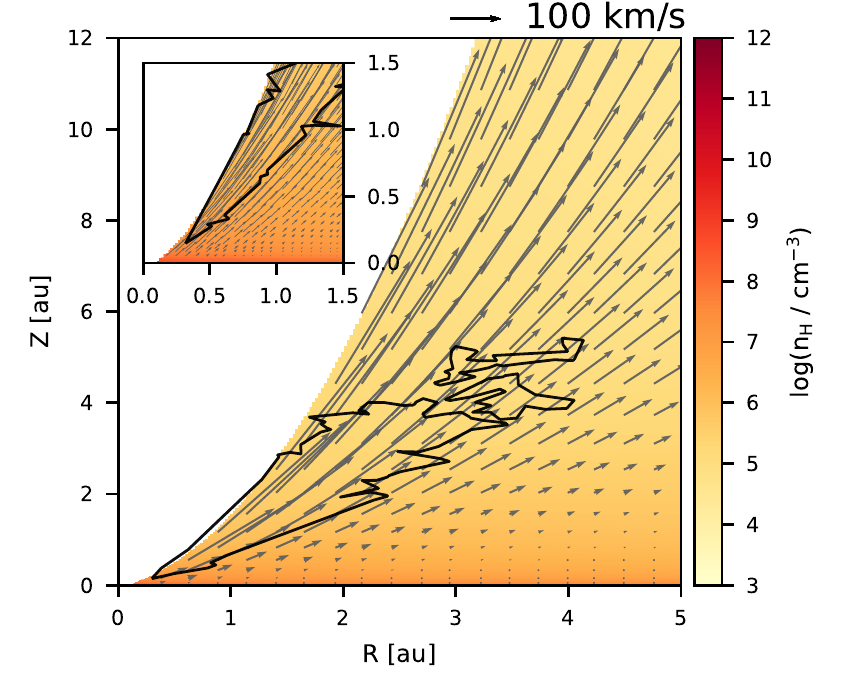}
\end{center}
\caption{Same as in Figure~\ref{fig:X-ray_density} but for the MHD wind model presented in Section~\ref{sec:MHD_model}. The inner region within 1.5 au $\times$ 1.5 au in the radial and vertical coordinates is shown in the overlaid panel on the top left corner. The superimposed contour lines define the 90$\%$ emission region for the free-free emission at 1 cm, after a 2D gaussian smoothing with $\sigma = 2$ grid cells to make the contour lines less noisy and confusing.
}
\label{fig:MHD_density}
\end{figure*}


\section{Results}
\label{sec:results}

Figures~\ref{fig:X-ray_density} and \ref{fig:MHD_density} present the density and velocity maps for the neutral hydrogen as derived from the simulations presented in Sections~\ref{sec:X-ray_model} and \ref{sec:MHD_model} for the photoevaporation and MHD wind models, respectively. The contour lines show the regions in the $(r,z)$ plane where the bulk of the emission at 1 cm is originated from.

The comparison between the X-ray photoevaporation and MHD wind models highlights significant differences between the spatial distribution of the emission. These differences are evident in Figure~\ref{fig:radial_profiles}, which shows the radial profiles for the surface brightness of the ionized gas emission at 1 cm for three models calculated in this work.

The photoevaporation disk models predict emission with two main spatial components. The first is localized at stellocentric distances of $\approx 0.5-0.9$ au, and corresponds to emission originated from the hot inner disk atmosphere, which is gravitationally bound to the star-disk system. 
The second component is spatially much broader, as it extends from about 1 to $4-5$ au from the star. The radial shape of this component  depends on the X-ray luminosity of the star, as a higher luminosity ($L_X = 10^{30}$ erg s$^{-1}$) produces significantly stronger emission, with a pronounced peak at a stellocentric distance of about 3 au (Fig.~\ref{fig:radial_profiles}, middle panel), than a model with lower X-ray luminosity ($L_X = 10^{29}$ erg s$^{-1}$, Fig.~\ref{fig:radial_profiles}, top panel). Contrary to the first component which comes from the bound disk atmosphere, this spatially broader component originates from ionized gas in the unbound wind.  

Compared to the disk photoevaporation model, the predicted radial profile for the MHD wind model is much smoother, and the emission extends to the very inner edge of the grid (Fig.~\ref{fig:radial_profiles}, bottom panel). This is a manifestation of the fact that whereas a photoevaporative thermal wind cannot be launched very close to the star, because of its intense gravity, a MHD wind can instead develop also from those regions.
We note here that the emission from the MHD wind model is originated only from the wind as a disk is not present in the analytical model presented in Section~\ref{sec:MHD_model}. The addition of a disk would not modify the results shown here as the wind would strongly screen the disk from ionising  radiation.

The 2D maps are presented in Figure~\ref{fig:model_1cm}, which shows both the synthetic model maps (left column) and the results of the simulated ngVLA observations (right column). We note here that the rather pronounced multiple narrow rings which are visible on the X-ray photoevaporation models, especially with $L_X = 10^{29}$ erg s$^{-1}$, are an effect of the noise in the model (see Appendix~\ref{sec:appendix}), and therefore these small-scale substructures in the emission are not physical. The total flux densities at 1 cm are 0.05, 0.15 and 1.8 mJy for the photoevaporation model with $L_X = 10^{29}$ and $10^{30}$ erg s$^{-1}$, and the MHD wind model, respectively. 

The maps on the right column show how the ngVLA, with its sensitivity and angular resolution of a few milliarcsec at 1 cm, would be able to resolve the main structures expected in the continuum emission at 1 cm for the photoevaporation and MHD wind models presented here for TW Hya. 

\section{Discussion}
\label{sec:discussion}

The results presented in the previous sections indicate that future radio observations at high angular resolution with an instrument like the ngVLA can spatially resolve sub-structures in the free-free gas emission expected by models of disk photoevaporation and MHD winds.  

\begin{figure*}[ht!]
\begin{center}
\includegraphics[scale=0.65]{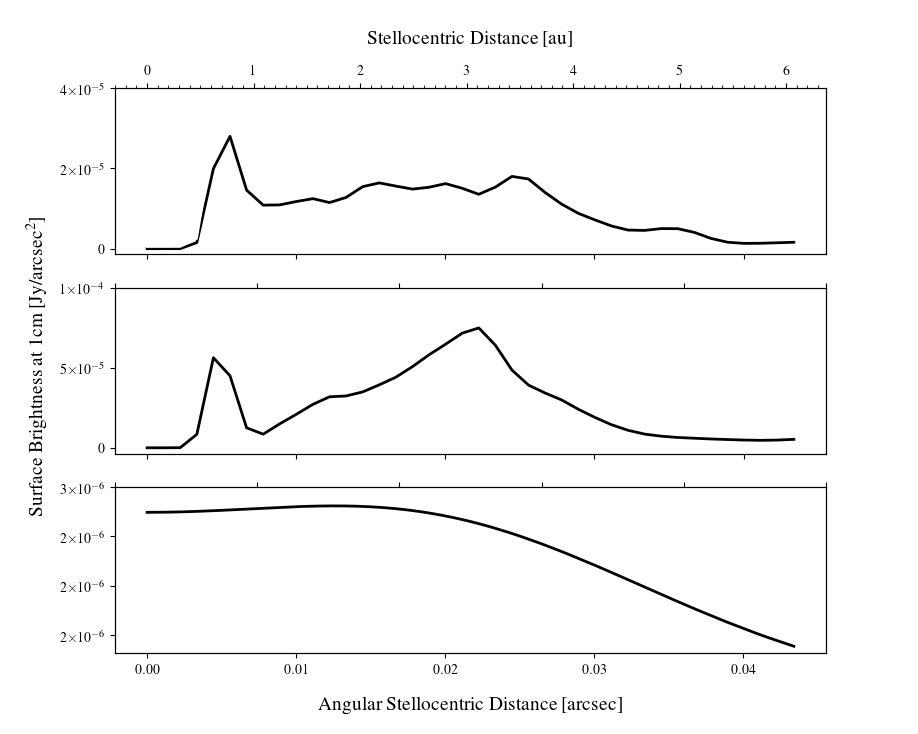}
\end{center}
\caption{\footnotesize {
Radial profiles of the surface brightness of the free-free emission at a wavelength of 1 cm for the three models calculated in this work. From top to bottom: X-ray photoevaporation model with $L_X = 10^{29}$ erg s$^{-1}$, X-ray photoevaporation model with $L_X = 10^{30}$ erg s$^{-1}$, MHD wind models.}}
\label{fig:radial_profiles}
\end{figure*}

The models of this study were adjusted to the specific case of TW Hyadrae, for which the process of disk photoevaporation has been proposed to explain the formation of the gap at 1 au from the star, together with other properties of the TW Hya system \citep{Ercolano:2017}. Figure~\ref{fig:radial_profiles} shows that disk photoevaporation models predict a bright ring at 1 cm very close to the location of the dust gap at sub-mm/mm wavelengths. Future radio observations can test this prediction. Also, multi-wavelength observations at cm wavelengths can be used to constrain the physical mechanism responsible for the emission by measuring the spectral index \citep[e.g.,][]{Rodmann:2006}.

Also the MHD wind model presented here and adjusted to the case of TW Hya predicts significant emission at 1 cm from the ionized gas in the wind. The hypotheses of photoevaporation-driven and MHD winds can be tested by constraining the spatial profile of the emission, as radiation from an MHD wind can extend to radii in the disk much closer to the star (Figure~\ref{fig:radial_profiles}).
This is in general agreement with the interpretation of the different components identified in the line profiles of typical wind diagnostics towards young stars \citep[e.g.,][]{Weber:2020}.

The results of our photoevaporation disk model with different values for the X-ray luminosity of the star show different radial profiles for the surface brightness at 1 cm, both in terms of the peak intensity and the radial shape of the emission. For all the models calculated in this work, the bulk of the emission is produced within about 5 au from the central star, corresponding to about 0.08 arcsec at the distance of TW Hya. However, the intensity of the first peak associated to emission from hot gas bound to the disk, as well as the spatial extent of the emission from the photoevaporative wind further from the star depends on the X-ray luminosity of the star itself. Given the estimate for the X-ray luminosity for the TW Hya star, we expect the radio emission to show two prominent peaks, at stellocentric distances of $\approx 0.5 - 1$ au and $3 - 3.5$ au, respectively. We note that the total flux from the model with $L_X = 10^{30}$ erg s$^{-1}$ is more in line with the fluxes measured at cm wavelengths, even though this system lacks strong constraints on the flux from ionized gas at 1 cm \citep[][]{Pascucci:2012}. The expected flux from our MHD wind model is a factor of $\sim 10\times$ higher than in the photoevaporation case. It is worth noticing that our assumed mass loss rate is comparable to some of the mass accretion rate estimates for the TW Hya young star in the literature \citep[e.g.,][]{Brickhouse:2012}. Although this is line with some previous predictions from models of disk winds which account for non-ideal MHD \citep[e.g.,][]{Bai:2013}, other similar models with different assumptions predict mass loss rates lower than mass accretion rates by factors of $\sim 2 - 10$ \citep[e.g.,][]{Gressel:2020}.

A broader investigation of the model parameters, not necessarily related to the specific case of TW Hya, would be necessary to better characterize the dependence of the radio-emission on the stellar X-ray luminosity, mass loss rate, and other parameters of the disk-star system. 
Such investigation would be important also for the interpretation of the radio emission already observed towards several young low-mass stars in nearby star forming regions \citep[e.g.,][]{Rodmann:2006,Macias:2016,Ubach:2017}, as well as to predict possible trends between some of the key physical quantities for the star and disk, which would guide future surveys of young stellar objects (YSOs) at radio wavelengths.

Although this work focuses on the case of TW Hya, which is a nearly face-on disk, high-angular resolution observations of disks in nearly edge-on orientations have the potential to resolve the vertical extent of the free-free emission from the wind. This would provide another way to test the predictions of X-ray photoevaporation and MHD wind models. Whereas computing results at different inclination angles is beyond the scope of this study, we plan on performing this investigation in a future work.

\begin{figure*}[ht!]
\begin{center}
\hspace*{-0.3cm}
\includegraphics[scale=0.8]{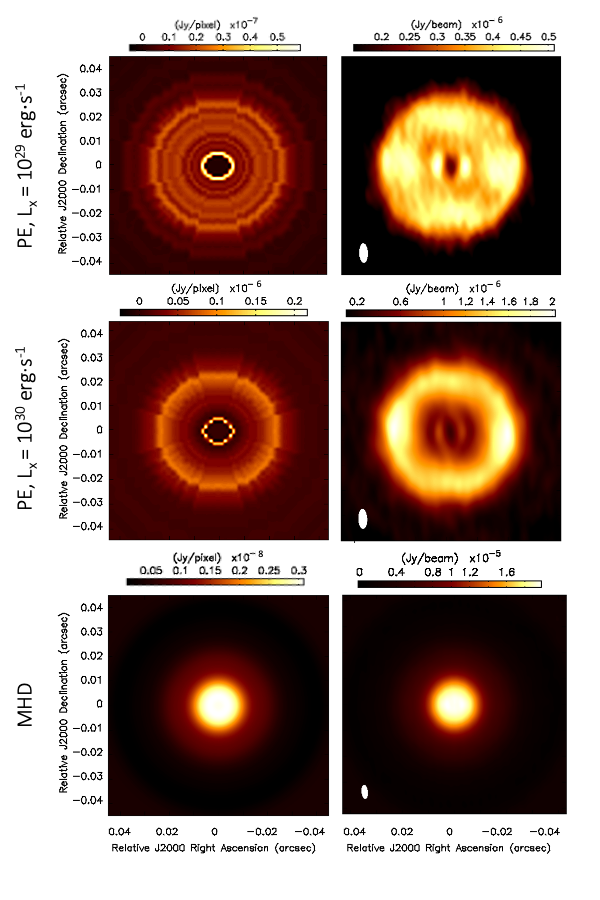}
\end{center}
\vspace{-5mm}
\caption{\footnotesize {Synthetic model images (left column) and ngVLA simulated observations (right column) for the ionized gas continuum emission at 1 cm. The top and middle panels show the maps from the photoevaporation disk models with X-ray stellar luminosities of $10^{29}$ and $10^{30}$ erg s$^{-1}$, respectively. The bottom panels show the maps from the MHD disk models. The unit for the surface brightness in each map of the models is Jy/pixel, where the sides of the square pixels have sizes of 0.8 and 0.2 mas for the photoevaporation and MHD models, respectively.
In the ngVLA maps the synthesized beam has sizes of $5.2~\rm{mas} \times 2.2~\rm{mas}$ with a position angle of 5.3 degrees. The rms noise on each ngVLA map is 27 nJy/beam.
Note that the intensity color scales on the maps are different.}}
\label{fig:model_1cm}
\end{figure*}

The disk models presented here neglect the contribution of the dust emission at the wavelength of 1 cm considered in this study. This is justified by an analysis of the dust continuum emission measured at 0.87 mm for the TW Hya disk and presented in \citet{Andrews:2016}, which shows a peak brightness temperature of 30 K. This can be considered as an upper limit for the brightness temperature at 1 cm under the assumption that the dust emission is optically thick even at cm wavelengths. The peak brightness would be only a factor of 1.5 higher than the rms noise in the maps presented in Fig.~\ref{fig:model_1cm} and in general a factor $> 10$ lower than the peak brightness temperature for the predicted free-free emission for the disk models presented in this work. Observations at wavelengths longer than 1 cm would further decrease the amount of contamination from dust thermal emission, but would have the critical disadvantage of providing poorer spatial resolution than at 1 cm. Since one of our main goals is to quantify the potential of future hi-res observations to test the model predictions for the spatial distribution of the disk free-free emission, we leave to a future work the investigation of the free-free emission at longer cm wavelengths.

\section{Conclusions}
\label{sec:conclusions}

We have presented predictions for the free-free emission at 1 cm from models of X-ray photoevaporation as well as of magnetocentrifugally driven MHD winds with parameters adjusted to the case of the TW Hya YSO. 

We showed that the X-ray photoevaporation models that have been proposed to explain the opening of the gap observed by ALMA in the dust thermal emission predict bright free-free emission in the region close to the gap in the TW Hya disk. This emission is characterized by two main spatial components: one at stellocentric distances of $\approx 0.5-1$ au, which is originated from hot ionized gas in the inner disk atmosphere close to the star; one broader component that extends to $\approx 4-5$ au from the star, which is associated to unbound gas in the photoevaporative wind. The intensity and radial slope of these components depend on the X-ray luminosity of the star.

The magnetocentrifugally driven MHD wind model predicts a spatially smoother free-free emission from ionized gas, which extends to the closest regions to the star in our model. Contrary to the case of disk photoevaporation where the closest regions to the star are bound to the stellar gravity, the MHD wind in our model is launched also from the innermost regions of the disk.    

The predicted integrated fluxes from our models at 1 cm are in line with those observed for TW Hya, as well as other YSOs in nearby star forming regions using current radio facilities such as the \textit{Karl G. Jansky Very Large Array} (VLA) and the \textit{Australia Telescope Compact Array} (ATCA). However, to spatially resolve the predicted free-free emission and potentially distinguish between the models presented here, observations providing angular resolutions below 10 mas at cm wavelengths are required.
We showed that a future ngVLA, with its current reference design, would have enough sensitivity and angular resolution to detect and spatially resolve the main structures predicted by these models.

The results outlined in this work refer to the specific case of TW Hya. 
A future investigation of the same models covering a broader region of the model parameter space would be necessary to explore the dependence of these predictions on the model parameters, and guide future observational surveys of protoplanetary disks at radio wavelengths. These have the potential to shed light on physical mechanisms which have been proposed to play a key role on the evolution and dispersal of protoplanetary disks, and therefore on the formation of planets. 



\vspace*{2mm}

\acknowledgements We thank the anonymous referee for his/her comments which helped to clarify the manuscript. This work was
supported in part by the ngVLA Community Studies program, coordinated by the National Radio Astronomy Observatory,
which is a facility of the National Science Foundation operated
under cooperative agreement by Associated Universities, Inc. BE acknowledges support from the DFG Research Unit "Transition Disks" (FOR 2634/1, ER 685/8-1). MW acknowledges support from the DFG Research Unit "Transition Disks" (FOR 2634/1, ER 685/11-1). This research
was partially supported by the Excellence Cluster ORIGINS, which is funded by
the Deutsche Forschungsgemeinschaft (DFG, German Research Foundation) under
Germany’s Excellence Strategy - EXC-2094 - 390783311. Part of the simulations
have been carried out on the computing facilities of the Computational
Center for Particle and Astrophysics (C2PAP).

\bibliographystyle{aasjournal}
\begin{singlespace}
\bibliography{vsi_bib}
\end{singlespace}

\appendix
\section{Improved photon packet statistics in low emission spectral regions}
\label{sec:appendix}

The standard version of the {\sc mocassin} code expresses the radiation field in term of photon packets, $\varepsilon(\nu)$, of constant energy, such that
\begin{equation}
    \varepsilon(\nu) = \varepsilon_0 = n\,h\,\nu,
\end{equation}
where $n$ is the number of photons contained in the packet and $\nu$ is the frequency of the photons (Lucy 1999). When a photon of energy $\varepsilon(\nu_a) = \varepsilon_0$ is absorbed, it is immediately re-emitted from the same location with frequency $\nu_e$, which is drawn stochastically from the probability density function, based on the local medium emissivities. The total energy of the packet remains the same, meaning that in practice the number of photons contained in the re-emitted packet is changed. Keeping the photon packet energy constant is a simple way of ensuring conservation of energy throughout the simulation domain and thus accelerate convergence (Lucy 1999).

\begin{figure*}[ht!]
\begin{center}
\includegraphics[scale=1.2]{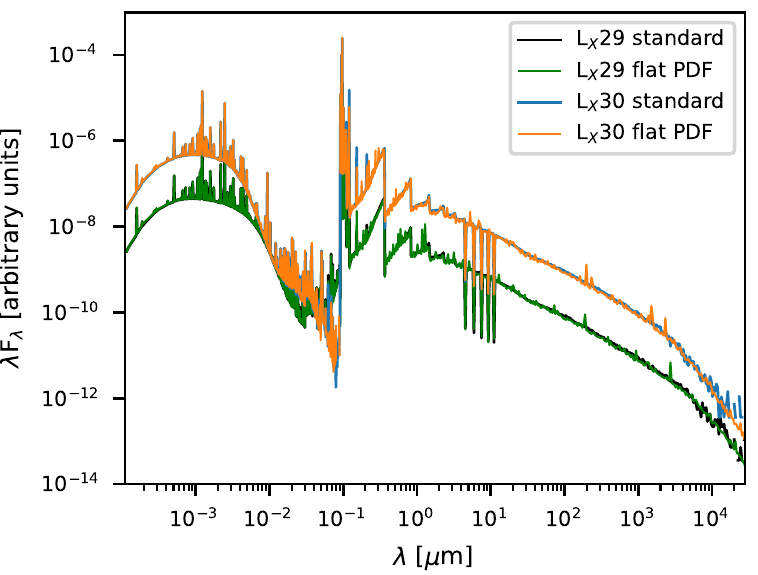}
\includegraphics[scale=1.2]{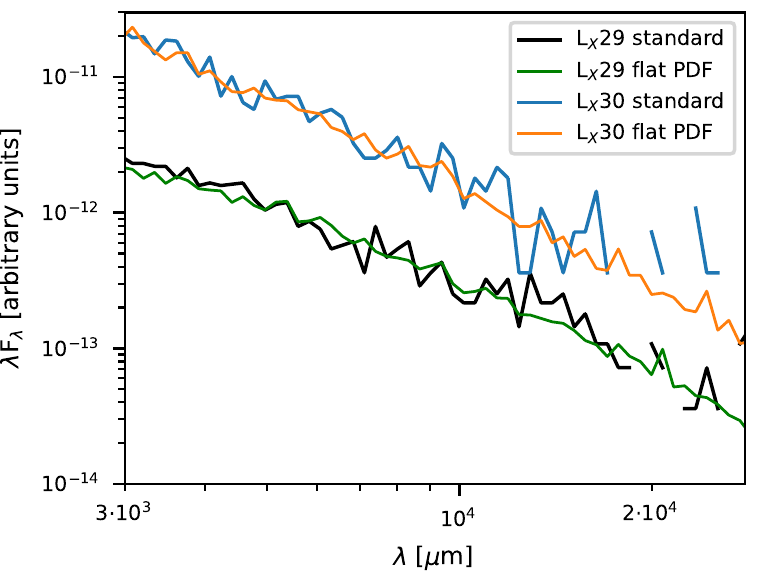}
\end{center}
\caption{Spectral energy distributions (SEDs) for the models shown in the paper integrated over all lines of sight. The left panel shows that the standard and flat PDF models agree on the global SEDs. The right panel shows that the flat PDF method is able to achieve a higher signal to (Monte Carlo) noise in the cm region, which is in the low energy tail of the SED.}
\label{fig:seds}
\end{figure*}

The probability density function, $p(\nu)$, of an energy distribution $f(\nu)$ is given by
\begin{equation}
    p(\nu) = \frac{f(\nu)}{\int f(\nu) d\nu}.
\end{equation}
 The disadvantage of stochastically casting packets of constant energy from $p(\nu)$ is that packets of frequencies   where $f(\nu)$ is small will have a very low probability of being cast. This means that the emerging spectral energy distribution will have low signal to (Monte Carlo) noise in those low $f(\nu)$ regions. 
 
 The cm-wavelength region in protoplanetary discs, which is dominated by free-free emission of the gas, is one example where the standard {\sc mocassin} approach will fail (see Figure~\ref{fig:seds}, blue and black lines). 
 An alternative approach, employed in this work, is to apply a "flat" probability density function, i.e. packets are assigned a random frequency when they are emitted, thus equally sampling all spectral regions. This approach effecively means that packets are being drawn from the following probability density function
 \begin{equation}
     p(\nu)_{\rm{flat}} = \frac{1}{\int d\nu}.
 \end{equation}
 
 In order to ensure energy conservation, the energy $\Delta E$ of the emitted packet must then be weighted by the local probability density function $p(\nu)$, such that
 \begin{equation}
     \Delta\,E = \Delta\,E_0 \frac{p(\nu) }{p(\nu)_{\rm{flat}}} = \Delta\,E_0 \frac{f(\nu)}{\int f(\nu)d\nu} \int d\nu.
 \end{equation}
 In the case of the primary radiation field, $\Delta\,E_0 = \frac{L_{\rm{source}}}{N_p}$, where $L_{\rm{source}}$ is the luminosity of the irradiating source and $N_p$ is the total number of energy packets to be employed in the simulation. In the case of the secondary radiation field $\Delta\,E_0$ is the energy of the absorbed energy packet. 
 The above ensures that the integral of all energy packets emerging from a given location still reflects the local emission spectrum, thus enabling conservation of energy at all locations.
 
 Figure~\ref{fig:seds} compares the total emerging spectral energy distribution (SED, integrated for all viewing angles) from our simulations using the standard {\sc mocassin} approach (black and blue lines) with the SED obtained using flat PDFs (orange and green lines). The top panel shows the entire simulated frequency range and demonstrates that the standard method and the flat PDF methods yield the same results in terms of the global emerging SED. The bottom panel shows a zoom of the same spectrum in the cm region, showing that the signal to (Monte Carlo) noise is much lower in the SED obtained using the flat PDF method. The SEDs were obtained using the same total number of energy packets, $N_p$.

\end{document}